\documentclass[12pt]{article}

\usepackage{setspace}

\usepackage{epsfig,latexsym,amsfonts,amsmath,amsthm,amssymb,amsbsy,multirow,slashed,wasysym,textcomp,dsfont,comment,mathtools,cancel,cite,diagbox,datetime,appendix,BOONDOX-calo}
\usepackage{tocloft}
\usepackage{bbold}
\usepackage{sidecap}
\usepackage{adjustbox}
\usepackage{pgfplots}
\usepackage{oplotsymbl}
\usepackage{tikzpagenodes}
\usepackage{adforn}
\usepackage{tikz}
\usetikzlibrary{backgrounds}
\usetikzlibrary{patterns}
\usetikzlibrary{arrows.meta}
\usetikzlibrary{shapes}
\tikzstyle{bag} = [align=center]
\usetikzlibrary{decorations.pathmorphing}
\usepackage[normalem]{ulem}
\usepackage[mathscr]{eucal}
\usepackage[version=3]{mhchem}
\def\({\left(}
\def\){\right)}
\def\[{\left[}
\def\]{\right]}

\usepackage{hyperref}
\usepackage{subcaption}

 \newcommand{\badat}{\begin{alignedat}}
 \newcommand{\eadat}{\end{alignedat}}
 \newcommand\scalemath[2]{\scalebox{#1}{\mbox{\ensuremath{\displaystyle #2}}}}
 \def\be{\begin{equation}}
\def\ee{\end{equation}}

\def\p{\partial}

\usepackage{color}

\newcommand{\pink}[1]{\textcolor{\pink}{#1}}

\definecolor{dblue}{rgb}{0.2,0.50,0.80}

\tikzset{
  branch cut/.style={
    decorate,decoration=snake,
    to path={
      (\tikztostart) -- (\tikztotarget) \tikztonodes
    },
    execute at begin to={{
      \draw[thick,green!60!black,-{Stealth[]}] ($(\tikztostart)!.8!-10:(\tikztotarget)$) -- ($(\tikztostart)!.8!10:(\tikztotarget)$) node[scale=.8,pos=.7,above left] {$\times (-1)$};
    }}
  }
}

\def\n{k}

\def\J{\mathcal{J}}

\def\O{\mathcal{O}}

\def\bh{{\bar h}}
\def\bz{{\bar z}}
\def\bw{{\bar w}}

\def\bh{{\bar h}}
\def\bz{{\bar z}}
\def\bw{{\bar w}}

\numberwithin{equation}{section} 
\pgfplotsset{compat=1.17} 
\begin{document}

 \begin{titlepage}
  \thispagestyle{empty}
  \begin{flushright}
  \end{flushright}
  \bigskip

  \begin{center}

                  \baselineskip=13pt {\LARGE \scshape{
               Revisiting the Shadow Stress Tensor \\[.5em]
               in Celestial CFT
                  }
         }

      \vskip1cm

   \centerline{  {Shamik Banerjee}${}^{1,2}$ and {Sabrina Pasterski}${}^{3}$ 

   }

\bigskip\bigskip
 \bigskip\bigskip

 \centerline{\em${}^1$  National Institute of Science Education and Research (NISER), Bhubaneswar 752050, Odisha, India}

 \vspace{.5em}

 \centerline{\em${}^2$   Homi Bhabha National Institute, Anushakti Nagar, Mumbai, India-400085
}

 \vspace{.5em}

 \centerline{\em${}^3$   Perimeter Institute for Theoretical Physics, Waterloo, ON N2L 2Y5, Canada
 }

\bigskip\bigskip

\end{center}

\begin{abstract}

We revisit the standard construction of the celestial stress tensor as a shadow of the subleading conformally soft graviton.  In its original formulation there is an obstruction to reproducing the expected $TT$ OPE in the double soft limit. We propose a modification to the definition which circumvents this obstruction and then extend this change of basis beyond the conformally soft and single helicity sectors. In the process we investigate how (non)-commutativity of double soft limits is tied to the decoupling of primary descendants, and how our choice of celestial basis determines which symmetries are manifest at the level of the OPE beyond the MHV sector. 

\end{abstract}

 \bigskip \bigskip \bigskip \bigskip

\end{titlepage}

\setcounter{tocdepth}{2}

\tableofcontents

\section{Introduction}

Over the past few years celestial holographers have made progress towards constructing a dual description for gravitational scattering in asymptotically flat spacetimes~\cite{Pasterski:2021raf}. Under this holographic map $\mathcal{S}$-matrix elements are mapped to correlators in a conformal theory living on the celestial sphere. The main advantage of this program is that it effectively reorganizes amplitudes in terms of the available symmetries~\cite{Strominger:2017zoo}.

Lorentz invariance guarantees that a basis of boost eigenstates will transform as quasi-primaries. In the simplest case of massless particles, we can reach boost eigenstates from energy eigenstates by performing a Mellin transform in the energies. The saddle point approximation tells us that when we push our Cauchy slice to null infinity to prepare the in and out states, this procedure picks out operators which are local on the celestial sphere but smeared along the generators of $\mathcal{I}^\pm$. However, from the bulk perspective there is some flexibility in our dictionary~\cite{Pasterski:2017kqt}. Namely, intertwiners like the shadow transform map quasi-primaries to quasi-primaries.

A major motivation for pursuing the Celestial CFT (CCFT) construction comes from the fact that the Lorentz symmetry is enhanced when we couple to gravity.   Intriguingly, a universal coupling to the subleading soft graviton~\cite{Cachazo:2014fwa} is equivalent to the Ward identity for a Virasoro symmetry~\cite{Kapec:2014opa} and gives rise to a candidate stress tensor~\cite{Kapec:2016jld}.  The fact that we have such a stress tensor hints that the dual might behave more like a local CFT than we would otherwise have grounds to assert. Moreover, this would na\"ively help fix the ambiguity in our basis: the standard $T{\cal O}$ OPE is reproduced by the single subleading soft graviton insertion, when the ${\cal O}$ operator is a `Mellin basis' operator.  However, $T$ itself is constructed as a shadow transform of the $\Delta=0$ conformally soft graviton.

So which basis should we be using?  When trying to phrase the soft physics in a standard 2D CFT language, it appears that the natural dictionary for currents~\cite{Kapec:2017gsg,Fotopoulos:2019vac,Fotopoulos:2020bqj,Kapec:2021eug} involves shadow transforms of the conformally soft modes~\cite{Donnay:2018neh,Pate:2019mfs,Guevara:2019ypd,Adamo:2019ipt,Puhm:2019zbl}.  However, we see that the Mellin basis is the natural dictionary for local operators capturing the finite energy radiative scattering states, with a continuous spectrum on the principal series. It has also been shown \cite{Banerjee:2020zlg,Banerjee:2021cly} that in the standard Mellin basis the subleading soft graviton theorem is equivalent to the Ward identity of $\widehat{\overline{sl}}_2$ current algebra. This has been checked explicitly by computing the celestial OPE of two positive helicity soft gravitons in the MHV sector and this computation shows that the OPE can be written as a linear combination of supertranslation and $\widehat{\overline{sl}}_2$ current algebra descendants of positive helicity gravitons.  Meanwhile, the construction of $w_{1+\infty}$ generators from the residues of the Mellin transformed amplitudes involves another intertwiner that appears in split signature: the light transform~\cite{Guevara:2021abz,Strominger:2021mtt}.

 While the saddle point approximation is strictly only valid for the radiative states, the standard procedure is to analytically continue off the principal series~\cite{Donnay:2018neh,Donnay:2022sdg} to study behavior of celestial amplitudes in the complex-$\Delta$ plane~\cite{Arkani-Hamed:2020gyp,Chang:2021wvv}.  In particular, we would like to be able to continuously take the conformally soft limits of the $T{\cal O}$ OPE. In doing so we encounter a variety of puzzles. Firstly, applying a shadow transformation only in this limit appears ad hoc. Secondly, we run into subtleties with double soft limits in either basis~\cite{Distler:2018rwu,Anupam:2018vyu,Fotopoulos:2019vac,Fotopoulos:2020bqj,Campiglia:2021bap,Kapec:2022hih}, which will be at the center of our explorations here.  Thirdly, even after resolving these issues, different basis choices make different symmetries manifest at the level of the OPE.

More succinctly: in a 2D CFT an operator and its shadow cannot both be treated as local operators.   In the celestial context we have an additional constraint: because the currents come from limits of the hard operators it seems like either our dictionary should have all of the operators be shadowed or not.  This presents a tension between our dictionary of local operators in the 2D theory, the existence of symmetry generators, and consistency of their algebra.\footnote{It would be interesting to explore how this generalizes to higher dimensions. Note that for $d>2$ we only need to demand that the operators are quasi-primaries which will still be the case in the shadow basis \cite{Kapec:2017gsg,Kapec:2021eug}. 
} Focusing on the subleading conformally soft graviton sector, we can phrase the puzzle as follows: We know that a positive helicity subleading soft graviton, after shadow transformation, becomes the \textit{antiholomorphic} stress tensor of a $2$D (celestial) CFT. So we expect that we should be able to write the OPE in the MHV sector as a linear combination of $\overline{Vir}$ primary and descendants. But, this is not what we find. Rather, the OPE is written in terms of supertranslation and $\widehat{\overline{sl_2}}$ current algebra descendants.  So a natural question is, what happens to the antiholomorphic stress tensor? In this paper, we provide an answer to this question.

We begin by revisiting the standard construction of the celestial stress tensor from~\cite{Kapec:2016jld}, defined as a shadow of the subleading soft graviton\cite{Cachazo:2014fwa,Kapec:2014opa}.   In its original formulation, there is an obstruction to reproducing the expected $TT$ OPE in the double soft limit considered in~\cite{Fotopoulos:2019vac}.
 With the insights of the celestial diamond and nested primary descendants~\cite{Banerjee:2019aoy,Banerjee:2019tam,Pasterski:2021fjn,Pasterski:2021dqe}
this can be cleanly phrased in terms of the weight $\Delta=-1$ reparameterization mode from which both the celestial stress tensor and the subleading soft graviton descend. Here, we propose a modification to the standard definition~\cite{Kapec:2016jld} which circumvents this obstruction and then extend this change of basis beyond the conformally soft and single helicity sector. This procedure is reminiscent of monodromy projections familiar from standard CFT in the context of extracting contributions from local operator exchanges to the conformal blocks~\cite{SimmonsDuffin:2012uy,Haehl:2019eae}. In the process, we show how the (non)-commutativity of double soft limits signal obstructions to the decoupling of primary descendants in correlators, as well as how our choice of basis determines which symmetries are manifest at the level of the OPE.

 This paper is organized as follows.  In section~\ref{sec:subsoft} we revisit double soft limits of the subleading soft graviton in the positive helicity sector, where we observe an obstruction to the standard stress tensor Ward identity. We can avoid this by introducing a modified shadow stress tensor. We then extend this modification beyond the conformally soft sector in section~\ref{sec:newshad} and to mixed helicity amplitudes in section~\ref{sec:mix}. Finally we conclude with a summary in section~\ref{sec:discuss}.
Some mathematical background is included in Appendix~\ref{app:gelfand}, while we make further contact with other incarnations of the celestial stress tensor in Appendix~\ref{app:cpx_sphere}, 
before showing more explicitly how our modified stress tensor avoids the aforementioned obstruction in Appendix~\ref{app:ope-mod}.

Before proceeding let us set up some notation.  Unless otherwise specified, we will consider massless scattering in (1,3) signature where the external momenta take the form
\be
p_i = \epsilon_i\omega_i(1+z_i\bz_i,z_i+\bz_i,i(\bz_i-z_i),1-z_i\bz_i),
\ee
where $\epsilon_i=\pm 1$ and $\omega_i\ge 0$. Upon performing a Mellin transform in the frequency variables $\omega_i$ the $\mathcal{S}$-matrix gets mapped to a correlator of operators
\be\label{mellindict}
\prod_{i=1}^n\int d\omega_i \omega_i^{\Delta_i-1} A(p_i)=\langle \phi^{\epsilon_1}_{h_1,\bh_1}(z_1,\bz_1)...\phi^{\epsilon_n}_{h_n,\bh_n}(z_n,\bz_n) \rangle,
\ee
which transform as quasi-primaries of weight $h_i=\frac{1}{2}(\Delta_i+J_i), \bh_i=\frac{1}{2}(\Delta_i-J_i)$ under the Lorentz group. Here $J_i$ matches the helicity of the $i^{th}$ particle in an `all out' convention.  In Euclidean CFTs, the shadow transform acts as an intertwiner between representations with Weyl reflected weights $h_i\mapsto 1-h_i,~~\bar{h}_i\mapsto1-\bar{h}_i$. We will use $\tilde{\phi}^{\epsilon_i}_{h_i,\bh_i}$ to denote the operators reached by composing the Mellin transform~\eqref{mellindict} with the 2D shadow transform. Unless necessary we will suppress the label $\epsilon_i$ on the external operators.

\section{Symmetries of the Subleading Soft Graviton}\label{sec:subsoft}
One feature of the celestial map~\eqref{mellindict} is that it sends powers of $\omega$ in the soft expansion of gauge bosons to poles in the conformal dimension at integer values of $\Delta$ when we analytically continue from the principal series capturing radiative states to the complex plane. We will focus on soft limits of the positive helicity graviton in this section.  The subleading soft graviton 
is picked out as the residue at $\Delta= 0$ of the spin  $h-\bar{h}=2$ operator $\phi_{h,\bar{h}}$ corresponding to the positive helicity gravitons
\be\label{Sdef}
S(z,\bz)=\mathrm{Res}_{\Delta=0}\phi_{h,\bar{h}}(z,\bz).
\ee
The soft theorem~\cite{Cachazo:2014fwa} tells us that this limit has a universal form in celestial correlators
\be\label{singles}
\langle{S(w,\bar w) \prod_{i=1}^n \phi_{h_i,\bar h_i}(z_i,\bar z_i)}\rangle = - \sum_{k=1}^n \frac{\(\bar z_k - \bar w\)^2 \bar\partial_k + 2\bar h_k \( \bar z_k - \bar w\)}{w- z_k}\langle{\prod_{i=1}^n \phi_{h_i,\bar h_i}(z_i,\bar z_i)}\rangle .
\ee
The operator $S$ has weights $(1,-1)$. If we take its shadow we land on a $(0,2)$ operator
\be
\label{shadow}
\bar T(z,\bar z) = \frac{3}{\pi} \int d^2w \frac{1}{\(\bar z - \bar w\)^4} S(w,\bar w),
\ee
which was proposed as a candidate stress tensor in~\cite{Kapec:2016jld}.  Indeed, so long as all of the other operators are hard one can show that
\be\label{singlet}
\langle{\bar T(z,\bar z) \prod_{i=1}^n \phi_{h_i,\bar h_i}(z_i,\bar z_i)}\rangle = \sum_k \( \frac{\bar h_k}{\(\bar z - \bar z_k\)^2} + \frac{1}{\bar z - \bar z_k} \frac{\partial}{\partial \bar z_k}\) \langle{\prod_{i=1}^n \phi_{h_i,\bar h_i}(z_i,\bar z_i)}\rangle
\ee
matching the expected conformal Ward identity. However, we must revisit our construction if we want to be able to consistently take a second particle soft. 

As discussed in~\cite{Lipstein:2015rxa,Klose:2015xoa} one expects the double soft limits of amplitudes in the single helicity sector to commute, while this is not the case for particles of opposite helicities.\footnote{This can happen for spin-0 excitations when the conformal manifold has nontrivial curvature. See~\cite{Kapec:2022axw} for a nice recent exploration of marginal deformations of celestial CFTs.} The issue that we run into here arises from the appearance of the shadow transformation in our definition of $\bar{T}$. We can phrase this problem in an intrinsically 2D language, and its resolution will be closely related to the monodromy projections one needs to perform to select the stress tensor vs shadow blocks~\cite{Haehl:2019eae} in ordinary, non-celestial, CFTs. Namely, equation~\eqref{shadow} implies the following relationship between primary\footnote{Here we are talking about primaries of the global conformal group.} descendants of $S$ and $\bar{T}$
\be\label{eq:bottomdiamond}
\partial\bar T(z,\bar z) = - \frac{1}{2} \bar\partial^3 S(z,\bar z).
\ee
From the single soft theorem we see that these descendants vanish away from other hard operator insertions, namely
\be
\label{anti}
\langle{\partial\bar T(z,\bar z)\cdots}\rangle = \text{Contact Terms}
\ee
and
\be
\label{viol}
\langle{\bar \partial^3 S(z,\bar z)\cdots}\rangle = \text{Contact Terms}.
\ee
We want to emphasize that~\eqref{eq:bottomdiamond} holds at the level of the wavefunctions~\cite{Cheung:2016iub,Pasterski:2017kqt,Banerjee:2019tam,Pasterski:2021fjn} so whenever expressions \eqref{anti} and~\eqref{viol} are valid, the contact terms in~\eqref{anti} and~\eqref{viol} are related by the factor of $-2$ in~\eqref{eq:bottomdiamond}.  What will break down, however, is whether the operator equations~\eqref{anti} and~\eqref{viol} continue to hold in the presence of other soft insertions. 

\paragraph{A symmetry-based argument}
Before going into further detail, let us give a simple symmetry-based argument for why \eqref{anti} and \eqref{viol} cannot hold simultaneously as operator equations. In other words, why we expect $\bar\partial^3 S(z,\bar z)=0$ to be violated in a 2D CFT with an \textit{antiholomorphic} stress tensor. The argument goes as follows: Assume that we can consistently make one of the positive helicity hard gravitons in \eqref{singlet} subleading soft. This then implies that $S$ is a $\overline{Vir}$ primary of weight $\bar h =-1$. While $\bar\partial^3 S$ is a primary of the global conformal group but it is \textit{not} a $\overline{Vir}$ primary. Therefore the equation $\bar\partial^3 S=0$ is not $\overline{Vir}$ invariant and cannot hold in our 2D CFT.\footnote{We can restore the condition of being a Virasoro primary at the expense of introducing the superrotation Goldstone mode via the Weyl covariant derivatives in~\cite{Barnich:2021dta,Donnay:2021wrk}.  This vacuum structure becomes important at loop level~\cite{Pasterski:2022djr,Donnay:2022hkf} but should not appear in our discussion of the amplitudes at tree level, where the $\Delta=2$ mode is observed to decouple. }

This observation has implications for the symmetries of scattering amplitudes and the celestial CFT. For example, one can interpret the subleading soft graviton theorem \eqref{singles} as the Ward identity for $\widehat{\overline{sl}}_2$ current algebra \cite{Banerjee:2020zlg} and $S(z,\bar z)$ as the generating function for the corresponding $\widehat{\overline{sl}}_2$ currents. Our observation, together with \eqref{eq:bottomdiamond}, leads to the conclusion that \vspace{1em}

\noindent \textit{If we admit both $S(z,\bar z)$ and its shadow $\bar T(z,\bar z)$ as local operators in the theory then both the $\widehat{\overline{sl}}_2$ current algebra and the $\overline{Vir}$ symmetries will be violated}.\vspace{1em}

 \noindent As a result, the whole $w_{1+\infty}$ tower of symmetries \cite{Strominger:2021mtt} will be also be absent from the celestial theory. We will discuss this and other consequences later in the paper.

\paragraph{Shadows and Double Soft Limits}
Let us now explicitly show that~\eqref{eq:bottomdiamond} implies an inconsistency of combining consecutive soft limits with the shadow operation and expecting a conservation law for tree level amplitudes.  Starting from the stress tensor insertion~\eqref{singlet}, if we now send the conformal dimension of one of the hard particles to $0$ we see that the form of the prefactor on the right hand side of~\eqref{singlet} implies there will be non-contact terms in the following level-3 descendant
\be\badat{3}\label{Sdec}
&\langle{\bar T(z,\bar z) \bar\p^3 S(w,\bw)\prod_{i=2}^n \phi_{h_i,\bar h_i}(z_i,\bar z_i)}\rangle \\
&= \( - \frac{24}{\(\bar z - \bar w\)^5} - \frac{12}{\(\bar z - \bar w\)^4} \frac{\partial}{\partial \bar w}\)\langle S(w,\bw) {\prod_{i=2}^n \phi_{h_i,\bar h_i}(z_i,\bar z_i)}\rangle+\text{Contact Terms},
\eadat\ee
 where the prefactor comes from computing
 \be \[\p_\bw^{3},\( \frac{-1}{\(\bar z - \bar w\)^2} +  \frac{1}{\bar z - \bar w} \frac{\partial}{\partial \bar w}\) \]=\( - \frac{24}{\(\bar z - \bar w\)^5} - \frac{12}{\(\bar z - \bar w\)^4} \frac{\partial}{\partial \bar w}\)
 \ee
and we have used $\bar{h}=-1$ for the operator $S$. We know from~\eqref{singles} that this correlator has support for $(w,\bw)$ away from other operator insertions. Namely the primary descendant $\bar\p^3 S$ no longer vanishes as an operator in the presence of a celestial stress tensor insertion.  Using our descendancy identity~\eqref{eq:bottomdiamond}
 we see that 
\be\badat{3}\label{TdecT}
&\langle{\bar T(z,\bar z)  \p \bar T(w,\bw)\prod_{i=2}^n \phi_{h_i,\bar h_i}(z_i,\bar z_i)}\rangle \\
&= \(  \frac{12}{\(\bar z - \bar w\)^5} + \frac{6}{\(\bar z - \bar w\)^4} \frac{\partial}{\partial \bar w}\)\langle S(w,\bw) {\prod_{i=2}^n \phi_{h_i,\bar h_i}(z_i,\bar z_i)}\rangle+\text{Contact Terms}.
\eadat\ee
The presence of this non-contact term serves as an obstruction to the OPE for $\bar T$ that would be expected from the BPZ construction. 
By contrast, repeating the same manipulations starting from~\eqref{singlet} we would run into no such obstruction to
\be\badat{3}\label{SdecS}
\langle{  S(z,\bar z) \bar\p^3 S(w,\bw)\prod_{i=2}^n \phi_{h_i,\bar h_i}(z_i,\bar z_i)}\rangle= \text{Contact Terms}
\eadat\ee
and similarly
\be\badat{3}\label{SdecT}
\langle{  S(z,\bar z) \p \bar T(w,\bw)\prod_{i=2}^n \phi_{h_i,\bar h_i}(z_i,\bar z_i)}\rangle= \text{Contact Terms}
\eadat\ee
consistent with what we expect from the fact that soft limits of same helicity gravitons commute and the descendancy relation~\eqref{eq:bottomdiamond}, though this will no longer be the case if we also include the opposite helicity soft sector. Equation~\eqref{SdecS} tells us we can construct a consistent $\widehat{\overline{sl}}_2$ current algebra from the positive helicity subleading soft graviton while equation~\eqref{TdecT} tells us that the shadow stress tensor $\bar{T}$ does not generate a consistent $\overline{Vir}$ symmetry as written, due to an obstruction that arises when more than one operator goes soft.

\paragraph{A Modified Stress Tensor}
In the remainder of this section we will show that we can provide a modification to our definition for the stress tensor so that we get a consistent $\overline{Vir}$ symmetry. 
To do so we will take inspiration from the `celestial diamond' framework of~\cite{Pasterski:2021dqe,Pasterski:2021fjn} and discussion of gauge redundancies amongst nested primary descendants in~\cite{Banerjee:2019aoy,Banerjee:2019tam}. Namely, we will start by lifting both $S$ and $\bar T$ to a weight $(0,-1)$ operator $\epsilon$ 
\be\label{topdiamond}
S=\p \epsilon,~~~ \bar T=-\frac{1}{2}\bar\p^3 \epsilon
\ee
whose correlators take the form
\be\label{epscorr}
\badat{3}
&\langle{\epsilon(z,\bar z) \prod_{i=1}^n \phi_{h_i,\bar h_i}(z_i,\bar z_i)}\rangle \\ 
&= - \sum_{k=1}^n \(\ln\mu^2 |z- z_k|^2\) \{\(\bar z_k - \bar z\)^2 \bar\partial_k + 2\bar h_k \( \bar z_k - \bar z\)\}\langle{\prod_{i=1}^n \phi_{h_i,\bar h_i}(z_i,\bar z_i)}\rangle,
\eadat
\ee
where $\mu$ is an IR regulator of the 2D model.  Note that the $\mu$-dependence drops out because of the global Ward identity
\be
\sum_{k=1}^n \{\(\bar z_k - \bar z\)^2 \bar\partial_k + 2\bar h_k \( \bar z_k - \bar z\)\}\langle{\prod_{i=1}^n \phi_{h_i,\bar h_i}(z_i,\bar z_i)}\rangle = 0.
\ee
One can see that~\eqref{epscorr} reproduces both~\eqref{singles} and~\eqref{singlet} upon taking the appropriate derivatives~\eqref{topdiamond}.  Moreover~\eqref{eq:bottomdiamond} follows automatically from~\eqref{topdiamond}. The operator $\epsilon$ has the same weights as a reparameterization mode. While we will postpone this interpretation for the time being, the manipulations surrounding monodromy projections and distinguishing between local operator versus shadow exchanges, as encountered in~\cite{Haehl:2019eae}, are relevant to this story.\footnote{In the celestial context there are naturally two symplectically paired `memory' and `Goldstone' modes. The diamond descending from $\epsilon$ measures the spin memory effect~\cite{Pasterski:2015tva}, while the paired superrotation Goldstone mode is the more natural `reparameterization' operator~\cite{Himwich:2019qmj,Ball:2019atb} and descends to its own diamond of SL$(2,\mathbb{C})$ primaries. Focusing on the memory modes, if we consider splitting the correlator~\eqref{epscorr} into two parts by replacing
\be
\log\mu^2|z-z_k|^2=\log\mu (z-z_k)+\log\mu (\bz-\bz_k),
\ee
we see that the presence of both terms guarantees the $S$ and its shadow $\bar{T}$ both descend from this top corner of the diamond. 
With either the first or second term on its own, we can essentially kill either the left or right corner of the celestial diamond.  This different projections can be phrased as gauging one of two transformations~\cite{Haehl:2019eae} 
\be\label{eq:Sproj}
[\epsilon(z,\bz)]_S=\epsilon(z,\bz)+\Lambda(\bz),~~~~
[\epsilon(z,\bz)]_{\bar{T}}=\epsilon(z,\bz)+\Lambda_0(z)+\Lambda_1(z)\bz+\Lambda_2(z)\bz^2.
\ee
However, the additional subtlety we encounter here is that doing such a monodromy projection of the correlator not only impacts our prescription for both the $\epsilon$ operator explicitly shown in~\eqref{epscorr} but also the hard operators whose conformally soft limits are related to $\epsilon$ by~\eqref{topdiamond}.}

In the celestial context we are also interested in understanding the correct external operators rather than just how to project their correlators onto `physical' exchanges when we decompose them into lower point amplitudes.  This lends itself to a slightly different phrasing of the problem. As we will discuss in the next section, being able to consistently take the conformally soft limit implies a similar restriction on what hard operators we can insert. We can choose a different Green's function so that
\be\label{Slift}
S=\p \epsilon_S,~~~ \langle  \bar\p^3 \epsilon_S(w,\bw)  \prod_{i=1}^n \phi_{h_i,\bh_i}(z_i,\bz_i) \rangle=\text{Contact Terms},
\ee
where the left hand equation holds away from other operator insertions and the right hand equation assumes all other operator insertions are hard. More concretely, assuming that we can analytically continue our expression for the soft theorem~\eqref{singles} off of the celestial sphere $z^*=\bz$, we can define following non-local operator 
\be\label{eq:eps}
\epsilon_S =\int_{z_0}^z d w S(w,\bz), 
\ee
where the open contour runs between a reference point, $z_0$, and $z$. Since $S$ has weight $h=1$ the integrand has the appropriate left-handed weight. The correlation function of $\epsilon_S$ is given by
\be\label{epscorr2}
\badat{3}
&\langle{\epsilon_{S}(z,\bar z) \prod_{i=1}^n \phi_{h_i,\bar h_i}(z_i,\bar z_i)}\rangle \\ 
&= - \sum_{k=1}^n \ln\mu \(z- z_k\) \{\(\bar z_k - \bar z\)^2 \bar\partial_k + 2\bar h_k \( \bar z_k - \bar z\)\}\langle{\prod_{i=1}^n \phi_{h_i,\bar h_i}(z_i,\bar z_i)}\rangle.
\eadat
\ee

Now defining $\bar{T}$ via the usual shadow definition~\eqref{shadow}, we can construct the following candidate for a modified stress tensor\footnote{Note that it is compatible to view our correction term as removing a contribution from the hard charge, a weight $\Delta=3$, $J=-1$ operator constructed from the matter fields
\be
\p \bar T=j_{3,-1}~~\Leftrightarrow~~ \p(\bar T-\p^{-1}j_{3,-1})=0,
\ee
so that the Ward identity enforces conservation of the modified stress tensor, identically~\cite{Hu:2022txx}.}
\be\label{Slift2}
\bar T_{\rm mod}= \bar T+\frac{1}{2}\bar\p^3\epsilon_S,~~~\p \bar T_{\rm mod}=0.
\ee
It's conservation follows from~\eqref{eq:bottomdiamond} and the first equation in~\eqref{Slift},  while the fact that it reduces to  $\bar{T}$ in the absence of other soft insertions follows from the second equation in~\eqref{Slift}.  If we repeat the same manipulations that led to the obstruction~\eqref{TdecT} with $\bar T$ replaced with $\bar T_{\rm mod}$, we indeed find that the non-contact terms vanish, as expected from the operator equation~\eqref{Slift}.\footnote{See appendix \ref{app:ope-mod} for the details.} However the analog of~\eqref{Sdec} implies that $\bar\p^3 S$ is non-vanishing in the presence of a $\bar T_{\rm mod}$ insertion so, much like the monodromy projection description, you give up the $\widehat{\overline{sl}}_2$ symmetry to get a manifest $\overline{Vir}$ symmetry.

\section{A Modified `Shadow Basis'}\label{sec:newshad}

 The essence of our construction of $T_{mod}$ boiled down to exploiting an ambiguity in the Green's functions lifting us from the subleading soft graviton to the reparameterization mode. This nested structure of primary descendants is known to persist for other conformally soft operators with (negative) integer weights. 
 While we now have a prescription for the stress tensor that reproduces the correct (centerless) Virasoro Ward identity for both single and double soft insertions, there is still a tension with the fact that all of the other operators are un-shadowed.  Even at the level of matching the single soft insertions, we are not shadow transforming all operators simultaneously, but constructing a highly non-local object only for conformally soft modes. In this section we discuss how to generalize our construction of $T_{mod}$ beyond the subleading conformally soft limit to general complex weights.

  Let's start with the integer point case first. We will focus on the holomorphic sector here, though an analogous discussion holds for the anti-holomorphic case. See also the discussion in Appendix~\eqref{app:gelfand}. Suppose $\phi_{h}(z)$ is a SL$(2,\mathbb{C})$ primary. Namely,
\be
L_0 \phi_h = h \phi_h, \  L_1 \phi_h = 0 .
\ee
The representation is spanned by the descendants $L_{-1}^n \phi_h$. A `primary descendant' will be a state in this module such that
\be
\psi_n = L_{-1}^n \phi_h,~~~
L_1 \psi_n = 0 .
\ee
Applying the standard commutation relations we get 
\be
L_1 \psi_n = n \( n + 2h -1\) L_{-1}^{n-1} \phi_h .
\ee
Therefore a non-trivial `primary descendant' exists for the unique value
\be\label{nh}
n = 1-2h.
\ee
Now, in terms of the fields, we can write
\be
\psi_n(z) = \partial^{1-2h} \phi_h(z).
\ee
Since $h$ takes on the special (half) integer value~\eqref{nh}, this is an ordinary derivative.  However, we can try to analytically continue this to an arbitrary complex number $h$.

\paragraph{An Integral Representation}
Let us consider the differential equation 
\be\label{gn}
\frac{d^n}{dz^n} g_n(z) = g(z).
\ee
 One solution to~\eqref{gn} can be written as
\be
g_n(z) = \frac{1}{\Gamma(n)} \int_{a}^z dw (z -w)^{n-1} g(w).
\label{greensfn}
\ee
Here $g_n(z)$ represents the $n$-fold \textit{integral} of the function $g(z)$. Formally we can invert this by taking $n\mapsto -n$ in which case the  $n$-fold \textit{derivative} of the function $g(z)$ is given by
\be
\frac{d^n}{d z^n} g(z) = \lim\limits_{\epsilon\rightarrow 0}\frac{1}{\Gamma(\epsilon-n)} \int_{a}^z dw (z -w)^{\epsilon-n-1} g(w).
\ee
Here $n$ is still an integer. Analytically continuing this to general $n = 1-2h$ we can formally write
\be\label{lt}
\partial^{1-2h} \phi_h(z) = \frac{1}{\Gamma(2h -1)} \int_{z_0}^z \frac{dw}{(z -w)^{2-2h}} \phi_h(w) \equiv \phi'_{1-h}(z_0,z).
\ee
The choice of a reference point $z_0$ explicitly breaks the expected global conformal covariance, since this is clearly no longer a local operator. It will be convenient to take our reference point $z_0\rightarrow \infty$. While the point at infinity is not preserved by the global conformal group, we do have some control over the behavior of the scattering amplitudes at large angular separation. Indeed take $z_0\rightarrow\infty$ keeping $z$ fixed we expect $\phi_h(z_0)\rightarrow \frac{1}{z_0^{2h}}$.  Meanwhile, 
\be\label{limz}
\frac{\partial}{\partial z_0} \phi'_{1-h}(z_0,z) = - \frac{1}{\Gamma(2h-1)} \frac{\phi_h(z_0)}{(z-z_0)^{2-2h}}
\ee
so the leading term in $z_0$ is given by 
\be\label{lim2}
\frac{\partial}{\partial z_0}\phi'_{1-h}(z_0,z) \sim \frac{1}{z_0^2} \rightarrow 0, \ z_0 \rightarrow\infty.
\ee
It is natural to call~\eqref{lt} an `incomplete light ray operator.'   We see that the integrand is the same (and is the holomorphic `half' of the shadow kernel). As compared to the ones appearing in~\cite{Atanasov:2021cje,Sharma:2021gcz,Guevara:2021tvr} it extends over half the range.

\paragraph{Celestial Diamond Redux} Now let us see how these appear in our construction of a modified shadow basis. We will need to restore the dependence on $\bz$.  Consider a primary field $\phi_{h,\bar h}$ and its shadow $\tilde{\phi}_{1-h,1-\bar h}$. For generic weights we will define our shadow transform as follows
\begin{equation}
\label{def:2dShadowTransform}
\badat{3}
 \widetilde{ \O_{h,\bar h}}(w,\bar w)= \frac{K_{h,\bar h}}{2\pi}  \int d^2w' \frac{ \O_{h,\bar h}(w',\bw')}{(w-w')^{2-2h}(\bw-\bw')^{2-2\bar{h}}}\, . \\
\eadat
\end{equation} 
The normalization constant $K_{h,\bar h}$ was chosen to be $K_{h,\bar h}=h+\bar h+|h-\bar h|-1$  in~\cite{Pasterski:2017kqt,Pasterski:2021fjn}  so that
$\widetilde{ \widetilde{ \O_{h,\bar h}}}=(-1)^{2(h-\bar{h})} \O_{h,\bar h}$, matching the conventions of~\cite{Simmons-Duffin:2012juh}.  Meanwhile $K_{h,\bar h}=\frac{\Gamma(2-2 \bar h)}{ \Gamma(2 h-1)}$ in~\cite{Osborn:2012vt}.  We can leave it unspecified for the time being, so as to keep track of which statements are independent of this convention.

To connect to our discussion above, we need to understand the limiting behavior near integer points. The identity $\p_z \frac{1}{\bar{z}}=\pi \delta^{(2)}(z)$ (matching the conventions of~\cite{Osborn:2012vt}) tells us that for special integer weights the shadow kernel becomes a Green's function~\cite{Pasterski:2021fjn}, namely
\begin{equation}
\label{shadow_Ia_Ib}
 \p_{\bw'}^{\bar \n} \frac{(\bw'-\bw)^{\bar{\n}-1}}{ (w'-w)^{\n+1}} 
   =\pi (\bar{\n}-1)! \frac{(-1)^{\n}}{\n!} \p_{w'}^\n\delta^{(2)}(w'-w) \,,
\end{equation}
and similarly for the holomorphic sector for $k$ and $\bar k$ flipped.
Meanwhile from Gelfand~\cite{gelvol1} we know that
 \be\label{gel}
 \left[\frac{z^\lambda \bz^\mu}{\Gamma\left(\frac{1}{2}(\mu+\lambda)+\frac{1}{2}|\mu-\lambda|+1\right)}\right]_{\overset{\lambda=-k-1}{\mu=-l-1}}=\pi \frac{(-1)^{k+l+j}j!}{k!l!}\delta^{(k,l)}(z,\bz),
 \ee
  where $j=\frac{1}{2}(\mu+\lambda)-\frac{1}{2}|\mu-\lambda|=\min\{k,l\}$. We will explore these integer point representations in more detail in Appendix~\ref{app:gelfand}; however, the essence of the celestial diamond story we need here is that the shadow kernel can be interpreted as either Green's function or differential operators taking us between conformal primaries with Weyl reflected weights~\cite{Pasterski:2021fjn,Pasterski:2021dqe} (see also~\cite{Donnay:2022sdg}).  We can carry these observations away from the integer points by treating the shadow and light transforms as pseudo-differential operators.

The pseudo-differential operators introduced above are such that as we limit to integer points we have the following descendancy relation
\be\label{pseduosh}
(-1)^{2h-1}\Gamma(2-2h) \p^{2h-1} \tilde{\phi}_{1-h,1-\bar h}=K_{h,\bar h} \Gamma(2\bar h-1)  \bar\p ^{1-2\bar h}\phi_{h,\bar h}
\ee
generalizing~\eqref{eq:bottomdiamond}. The analog of~\eqref{Slift} is to introduce a field $f_{1-h,\bar h}$ such that
\be\scalemath{1}{\label{phicontact}
\phi_{h,\bar h}=(-1)^{2h-1}\Gamma(2-2h)\p^{2h-1}f_{1-h,\bar h},~~~
\langle  \bar\p^{1-2\bar h}f_{1-h,\bar h} 
\prod_{i=1}^n \phi_{h_i,\bh_i}(z_i,\bz_i) \rangle=\text{Contact Terms}}
\ee
so that the analog of~\eqref{Slift2} is
\be
\tilde{\phi}_{1-h,1-\bar h;mod}=\tilde{\phi}_{1-h,1-\bar h}-K_{h,\bar h}\Gamma(2\bar h-1)\bar\p ^{1-2\bar h}f_{1-h,\bar h},
\ee
while~\eqref{pseduosh} guarantees the analog of the conservation law $\p \bar T_{mod}=0$, implying the primary descendants that appear at integer points decouple. In terms of our explicit integral expressions for these pseudo differential operators we have 
\be\scalemath{.98}{\label{phimod}
\tilde{\phi}_{1-h,1-\bar h;mod}=\frac{K_{h,\bar h}}{2\pi}\left[\int_{\hat{\mathbb{C}}}d^2 w + \frac{i}{2}
(1-e^{2\pi i(2-2h)})\int_{-\infty}^z dw\int_{-\infty}^\bz d\bw \right]\frac{{\phi}_{h,\bar h}(w,\bw)}{(z-w)^{2-2h}(\bz-\bw)^{2-2\bh}} },
\ee
where
\be
K_{h,\bar h} = \frac{\Gamma(2-2\bar h)}{\Gamma(2h-1)} = \frac{\Gamma(2- 2h)}{\Gamma(2\bar h -1 )}, ~~~  h-\bar h = \pm 2.
\ee
The formula is the same for both positive and negative helicity gravitons.\footnote{For $\Delta=1-|J|-n$ for $n\in\mathbb{Z}_>$ we expect both holomorphic and antiholomorphic primary descendants from the representation theory, however we see that this would appear to involve an additional subtraction as compared to~\eqref{phimod}.} 

We see that if we complexify the shadow integral kernel, for generic weights there is a branch cut that extends from $w=z$ to $w=\infty$ in the complex $w$ plane and similarly, from $\bw=\bz$ to $\bw=\infty$ in the complex $\bw$ plane.   The monodromy around $z=w$ is the phase $e^{2\pi i(2h-2)}$.  Taking this into account, we can recognize the normalization as arising from the discontinuity across the branch cut in the complex $w$ plane for a particular orientation of the contour. We won't pursue this interpretation further here, but will revisit the complexification of the celestial sphere to $\mathbb{CP}^1\times \mathbb{CP}^1$ in appendix~\ref{app:cpx_sphere} in the context of comparing our subtraction to other constructions in the literature.

  Now we've seen that by construction $\p^{2h-1}\tilde\phi_{mod}$ vanishes in correlators. Because of~\eqref{phimod} and~\eqref{pseduosh} this is true both for contact terms as well so we are free to smear our correlators. Rather than restrict to the other operators being of $\phi_{h,\bh}$ type, they can also be shadowed or modified shadow operators
\be\label{conservedphimod}
\langle  \p^{2h-1} \tilde{\phi}_{1-h,1-\bar h;mod} \tilde{\phi}_{1-h_i,1-\bar h_i;mod}   \ldots \rangle={\rm Contact~Terms}.
\ee
In particular, for $h=1$ and $\bar h=-1$ this reduces to the conservation law for $\bar T_{mod}$ in these modified correlators
\be
\langle \p \bar T_{mod} \tilde{\phi}_{1-h_i,1-\bar h_i;mod}   \ldots \rangle={\rm Contact~Terms}.
\ee
In these expressions we've assumed that the $...$ included no other soft limits of the opposite helicity sector. We will now turn to the generic mixed helicity case.

\section{Mixed Helicity Amplitudes: A Proposal}\label{sec:mix}

Let us now go beyond the single helicity sector. The conventional double soft theorems for positive and negative helicity gravitons are known to have ambiguities~\cite{Klose:2015xoa}. Here we will see that sequential conformally soft limits present the same type of obstruction to the decoupling of primary descendants.  Again we will start by focusing on the subleading soft graviton.  Let $\bar S$ denote the $\Delta=0,J=-2$ soft graviton extracted as in~\eqref{Sdef} above, but for the negative helicity sector.  If we consider a mixed helicity correlator and take a sequential conformal soft limit where the $+$ helicity operator is taken to $\Delta=0$ first, we find
\be
\label{plus}
\begin{gathered}
\langle{S(w_1, \bar w_1) \bar S (w_2,\bar w_2) \prod_i \phi_i(z_i,\bar z_i)}\rangle_{- +} \\
= - \frac{\(\bar w_2 - \bar w_1\)^2 \bar\partial_{w_2} + 2\( \bar w_2 - \bar w_1\)}{w_1 - w_2} \langle{\bar S(w_2,\bar w_2) \prod_i \phi_i(z_i,\bar z_i)}\rangle \\
 - \sum_i \frac{\(\bar z_i - \bar w_1\)^2 \bar\partial_{i} + 2 \bar h_i \( \bar z_i - \bar w_1\)}{w_1 - z_i} \langle{\bar S(w_2,\bar w_2) \prod_i \phi_i(z_i,\bar z_i)}\rangle,
\end{gathered}
\ee
while taking the opposite order gives 
\be
\label{minus}
\begin{gathered}
\langle{S(w_1, \bar w_1) \bar S(w_2,\bar w_2) \prod_i \phi_i(z_i,\bar z_i)}\rangle_{+-} \\
= - \frac{\(w_1 -  w_2\)^2 \partial_{w_1} + 2\( w_1 - w_2\)}{\bar w_2 - \bar w_1} \langle{S(w_1,\bar w_1) \prod_i \phi_i(z_i,\bar z_i)}\rangle \\
 - \sum_i \frac{\( z_i -  w_2\)^2 \partial_{i} + 2 h_i \( z_i - \bar w_2\)}{\bar w_2 - \bar z_i} \langle{S(w_1,\bar w_1) \prod_i \phi_i(z_i,\bar z_i)}\rangle.
\end{gathered}
\ee
While one can further plug in the form of the soft theorem to evaluate the remaining correlators in these expressions and see that $(\ref{plus}) \ne (\ref{minus})$, we've paused here because we already encounter an obstruction to the null state decoupling we'd expect.  Namely, from the single soft limits we expect the primary descendants $\bar\p^3 S$ and $\p^3\bar S$ to decouple. However we see that in \eqref{plus} we have $\bar\partial^3 S (z,\bar z) = 0$ but $\partial^3 \bar S(z,\bar z) \ne 0$, while the opposite is true for \eqref{minus}.

Any attempt to try to evade the ambiguity by specifying a priori which order of soft limits to take still comes at a cost.  In the $(- +)$ case we preserve the $\widehat{\overline{sl}}_2$ current algebra but lose the $\widehat{sl}_2$ current algebra.  Similarly in the $(+ -)$ case the $\widehat{sl}_2$ current algebra still holds but we lose the $\widehat{\overline{sl}}_2$ symmetry. The fact that we are unable to preserve both decoupling conditions at the same time is perhaps not as surprising given that, taken together, the two current algebras do not form a closed algebra. Meanwhile we don't have obvious candidates for the additional generators of Diff$(S^2)$ (see also~\cite{Schwarz:2022dqf}). Phrased in this manner, we recognize this as the same type of problem that we started with in the single helicity sector. Namely, just as we saw from the double soft theorems that we couldn't consistently realize the $\overline{Vir}$ and the $\widehat{\overline{sl}}_2$ current algebra symmetries at the same, we similarly need to look for additional generators since the $\widehat{\overline{sl}}_2$ current algebra and $\overline{Vir}$ do not form a closed algebra.

By contrast the $\overline{Vir}$ and $\widehat{sl}_2$ generators form a closed algebra.  The same is true for $Vir$ and $\widehat{\overline{sl}}_2$ (and also for $Vir\times \overline{Vir}$). Can we use this to augment our soft algebras beyond the single helicity sector?  Let's return for a moment to the expression for the $(- +)$ limit given by~\eqref{plus}, and now do a shadow transformation of $S(w_1,\bar w_1)$
\be\label{res}
\begin{gathered}
\langle{\bar T(w_1,\bar w_1) \bar S(w_2,\bar w_2) \prod_i \phi_i(z_i,\bar z_i)}\rangle_{- +} \\
= \( \frac{1}{\(\bar w_1 - \bar w_2\)^2} + \frac{1}{\bar w_1 - \bar w_2} \frac{\partial}{\partial \bar w_2}\) \langle{\bar S(w_2,\bar w_2) \prod_i \phi_i(z_i,\bar z_i)}\rangle \\
+ \sum_i \( \frac{\bar h_i}{\(\bar w_1 - \bar z_i\)^2} + \frac{1}{\bar w_1 - \bar z_i} \frac{\partial}{\partial \bar z_i}\) \langle{\bar S(w_2,\bar w_2) \prod_i \phi_i(z_i,\bar z_i)}\rangle.
\end{gathered}
\ee
We see that the problem we encountered above goes away, namely $\p^3\bar S=0$ and $\p\bar T=0$.  One can check that the same is true if we started with the opposite choice $(+-)$ of conformally soft limits. Now of course from the previous section we know that we want to promote $\bar T\mapsto\bar T_{mod}$ to get a consistent $\overline{Vir}$ algebra but those statements were within that single helicity sector.

Taking a step back, we have seen that consecutive soft limits in the mixed helicity sector run into a problem because of the fact that we have both holomorphic and antiholomorphic poles. Equation \eqref{res} suggests that one way to resolve this issue will be to consider, instead of the standard~\cite{Pasterski:2017ylz} celestial amplitudes, amplitudes of the form
\be\label{1}
\langle{+ + + \ldots \widetilde -_{mod} \widetilde -_{mod}\widetilde -_{mod}\ldots}\rangle
\ee
and
\be\label{mpsh}
\langle{- - - \ldots \widetilde +_{mod} \widetilde +_{mod}\widetilde +_{mod}\ldots}\rangle
\ee
where one helicity sector has been mod-shadowed before taking any soft limit. 

In the first case, we expect that the symmetry that is realized locally is at least as big as the semi-direct product of  ${Vir}$ and $w_{1+\infty}$ and, similarly, in the second case it is the semi-direct product of ${\overline{Vir}}$ and $\overline{w}_{1+\infty}$. In this paper, while we will not try to give a proof of this proposal, we end this section by pointing out some of its novel features.  

The main point of our proposal is that the symmetry that is locally realized at the level of $\cal S$-matrix depends on the choice of basis for the asymptotic states. This may not be very surprising given that the basis transformation which takes us from \eqref{1} to \eqref{mpsh} is non-local. In one description we declare the mod-shadowed operators $\widetilde +_{mod}$ as local and in the other description, the conventional $+$ operators are local. But, there is \textit{no} useful description in which both the $+$ and $\widetilde +_{mod}$ coexist as local operators. This is consistent with our observation in section-$2$ that if we admit both $S(z,\bar z)$ and $\bar T(z,\bar z) = \tilde{S}(z,\bar z)$ as local operators in the theory then we lose both the $\widehat{\overline{sl_2}}$ and $\overline{Vir}$ symmetries. 

A relatively simple check of our proposal can be made in the MHV sector. First of all, it was shown in \cite{Banerjee:2021dlm} that the MHV sector has a symmetry group which is the semidirect product of $Vir$ and $w_{1+\infty}$. This is consistent with our proposal here. Furthermore, we know that when the MHV sector is described by the standard amplitudes $\langle{- - + + + \cdots +}\rangle$, the celestial OPE of two gravitons can be written \cite{Banerjee:2020zlg} in terms of supertranslation and $\widehat{\overline{sl}}_2$ current algebra descendants. According to our proposal if instead, we describe the MHV sector by amplitudes of the form $\langle{- -  \widetilde +_{mod} \widetilde +_{mod}\cdots\widetilde +_{mod}}\rangle$ then one should be able to write the celestial OPE between two gravitons as a linear combination of supertranslation and $\overline{Vir}$ descendants. This will be a nontrivial check of our proposal and we hope to return to some of these problems in near future.

\section{Discussion}\label{sec:discuss}
Since the early days of celestial CFT, there has been an underlying debate about which scattering basis in the bulk should correspond to the local operators in the celestial dictionary.  From the point of view of the global Lorentz symmetry any intertwiner between primaries should give an equivalent basis.  However, the extrapolate dictionary~\cite{Pasterski:2021dqe} seems to prefer the Mellin transformed basis because of the way that the momentum space and position space celestial spheres get identified under the large-$r$ saddle point near null infinity~\cite{He:2014laa}. The enhancement from global SL$(2,\mathbb{C})$ to a Virasoro symmetry would seem to settle this debate if it were not for the fact that the symmetry generators are constructed from a non-local shadow transform, while their Ward identities transforming hard operators demand that those are defined in terms of the ordinary Mellin transformed basis.

Here we've shown that there is a tension with the definition of the stress tensor as the shadow operator and shown how to remedy it.  The underlying reason for this tension is more general than our CCFT application: an operator and its shadow can't both be local.  However the fact that we are necessarily confronted with it is unique to the celestial case. We cannot separately prescribe different dictionaries for the hard and soft modes without being able to interpolate between the two when we continuously deform our hard operator dimensions to various conformally soft limits. In particular, the Virasoro symmetry picks out local operator for hard particles but wants to have its cake and eat it to with the soft graviton mode defining the stress tensor.

Our modification involves a certain projection of one half of the celestial diamonds so that there is only one radiative soft mode.  We then showed how to extend this to operators of any weight, so that we can go to a basis where the corrected shadow transform is again a conformally soft limit. In the end, we've seen that different bases make different symmetries manifest at the level of the celestial OPE and explored how certain helicity-asymmetric representations of the amplitudes can be better suited to taking multi-soft limits.  These explorations expand upon a variety of exciting approaches to elucidate the symmetries and structure of the 4D $\mathcal{S}$-matrix~\cite{He:2014cra,Banerjee:2020zlg,Fotopoulos:2019vac,Guevara:2021abz,Strominger:2021mtt,Sharma:2021gcz,Donnay:2021wrk,Donnay:2022sdg,Schwarz:2022dqf,Banerjee:2020vnt}.

\subsection*{Acknowledgements}
\vspace{-1mm}

We would like to thank  Yangrui Hu, Ashoke Sen, Atul Sharma, Andrew Strominger, Tomasz Taylor and Herman Verlinde for useful conversations. The work of SB is partially supported by the  Swarnajayanti Fellowship (File No- SB/SJF/2021-22/14) of the Department of Science and Technology and SERB, India and by SERB grant MTR/2019/000937 (Soft-Theorems, S-matrix and Flat-Space Holography).   SP is supported by the Celestial Holography Initiative at the Perimeter Institute for Theoretical Physics and has been supported by the Sam B. Treiman Fellowship at the Princeton Center for Theoretical Science. Research at the Perimeter Institute is supported by the Government of Canada through the Department of Innovation, Science and Industry Canada and by the Province of Ontario through the Ministry of Colleges and Universities.

\appendix

\section{Lessons from Gelfand}\label{app:gelfand} 
In this appendix we briefly summarize some results from volumes 1 and 5 of Gelfand's series on generalized functions~\cite{gelvol1,gelvol5}.   In particular, we will want to  review his construction of homogeneous generalized functions and the behavior of our SL$(2,\mathbb{C})$ primaries near integer conformal dimensions so that we can understand various contact terms that appear and how to handle their analytic continuations to other signatures.

\subsection{Generalized Homogeneous Functions}

Appendix B of Gelfand Volume 1~\cite{gelvol1} discusses generalized functions of a complex variable.  These are relevant for our understanding of the distributions that appear when we take descendants at special values of the conformal dimension. In particular our mode expansion of the conformal primaries involves so-called homogeneous generalized functions, whose definitions and properties we review here.  

Following~\cite{gelvol1}, a function $F(z,\bz)$ is a {\it homogeneous function of degree $(\lambda,\mu)$} if for $a\in \mathbb{C}-\{0\}$
\be
F(az,\bar a \bz)=a^\lambda \bar a^\mu F(z,\bz).
\ee
This can be promoted to a generalized function by considering the following functional
\be\label{pairing}
(F,\varphi)=\int d^2 z F(z,\bz)\varphi(z,\bz)
\ee
for an appropriate set of test functions $\varphi$ with bounded support so that we are free to integrate by parts
\be
(f,\varphi^{(j,k)})=(-1)^{j+k}(f^{(j,k)},\varphi).
\ee
In particular, via a change of variables we see that for our homogeneous functions we have
\be
(F,\varphi(z/a,\bz/\bar a))=a^{\lambda+1} \bar a^{\mu+1}(F,\varphi).
\ee
For ${\rm Re}(\mu+\lambda)>-2$ we can defined the generalized function $z^\lambda \bz^\mu$ via the functional $(z^\lambda \bz^\mu,\varphi)$ which converges for $\varphi$ infinitely differentiable with bounded support.  To extend this to ${\rm Re}(\mu+\lambda)<-2$ we need to regulate the integral. Gelfand identifies a prescription such that the answer away from {\it integer points} (see below) is defined by analytic continuation in the scaling dimensions from the region ${\rm Re}(\mu+\lambda)>-2$ where the integral converges. This function is analytic away from $\lambda,\mu=-k-1$ for $k\in \mathbb{Z}_>$ where it has simple poles  
\be
{\rm res}_{\overset{\lambda=-k-1}{\mu=-l-1}}(z^\lambda \bz^\mu,\varphi)=\frac{2\pi}{k!l!}\varphi^{(k,l)}(0,0).
\ee
Here the residue is in the single complex variable $s=\mu+\lambda$ with $n=\mu-\lambda\in\mathbb{Z}$ fixed.  Since the difference in scaling dimensions is integer valued both $\mu$ and $\lambda$ will be integer valued at these points, and are naturally labeled by $k,l\in\mathbb{Z}_>$ as above.

We can formally attach this residue of the integral to the generalized function $z^\lambda\bz^\mu$
\be
{\rm res}_{\overset{\lambda=-k-1}{\mu=-l-1}}z^\lambda \bz^\mu=\frac{2\pi}{k!l!}(-1)^{k+l}\delta^{(k,l)}(0,0).
\ee
 By introducing a gamma function with the same pole locations we can define an object whose limits to integral values is the distribution
 \be
 \left[\frac{z^\lambda \bz^\mu}{\Gamma\left(\frac{1}{2}(\mu+\lambda)+\frac{1}{2}|\mu-\lambda|+1\right)}\right]_{\overset{\lambda=-k-1}{\mu=-l-1}}=\pi \frac{(-1)^{k+l+j}j!}{k!l!}\delta^{(k,l)}(z,\bz),
 \ee
 where $j=\frac{1}{2}(\mu+\lambda)-\frac{1}{2}|\mu-\lambda|=\min\{k,l\}$. In particular, this implies that 
\be\label{d2z}
\p_\bz z^{-1+\alpha}\bz^{\alpha}=\alpha z^{-1+\alpha}\bz^{-1+\alpha}~~\Rightarrow~~\lim\limits_{\alpha\rightarrow 0} \p_\bz z^{-1+\alpha}\bz^{\alpha}=\pi \delta 
(z,\bz).
\ee
 This replaces the familiar relation $\p_\bz z^{-1}=\pi\delta(z,\bz)$ with a form that we can more readily analytically continue between signatures. Specifically, we see that the definitions of these generalized functions are implicitly tied to our integration contour in~\eqref{pairing} (here the Riemann sphere).

\subsection{Representations at Integer Points}

Volume 5 of Gelfand~\cite{gelvol5} discusses infinite dimensional representations of the Lorentz group. Each such representation is labeled by a pair of complex numbers $\chi=(n_1,n_2)$ where  $\(n_1 - n_2\)\in \mathbb Z$. The representation operator $T_{\chi}(g)$, for $g\in SL(2,\mathbb C)$, acts on the space $D_{\chi}$ of functions $\phi(z,\bar z)$ via
\be
T_{\chi}(g) \phi(z,\bar z) = (\beta z + \delta)^{n_1 -1} (\bar\beta \bar z + \bar\delta)^{n_2 -1} \phi\( \frac{\alpha z + \gamma}{\beta z + \delta},\frac{\bar \alpha \bz + \bar\gamma}{\bar \beta \bz + \bar \delta}\).
\ee
We see that that in terms of our usual notation for the weights $(h,\bar h)$ we have
\be
n_1 = 1 - 2h, \   n_2 = 1- 2\bar h, \  n_1 - n_2 = -2 (h - \bar h) = -2 J \in \mathbb Z.
\ee
The shadow transform induces a simultaneous Weyl reflection on the left and right handed weights. This corresponds to the map ${\rm Sh} : D_{\chi}\rightarrow D_{-\chi}$. Namely $\chi = (n_1, n_2) \rightarrow - \chi = (-n_1, - n_2)$ while
\be
{\rm Sh}\circ \phi(z,\bar z) = \int d^2z_1 (z - z_1)^{-n_1 -1} (\bar z - \bar z_1)^{-n_2 -1} \phi(z_1,\bar z_1).
\ee
Now suppose $\chi = (h,\bar h)$ is a representation where $h$ and $\bar h$ are neither simultaneously positive nor simultaneously negative integers. Then the shadow transform is one-to-one and onto
\be
\chi = (n_1,n_2) \sim -\chi = (-n_1,-n_2).
\ee

The so-called {\it integer points} are special representations for which $\chi = (n_1,n_2)$ are either simultaneously positive or simultaneously negative.  These are at the heart of the null state relations and celestial diamonds of~\cite{Banerjee:2019aoy,Banerjee:2019tam,Pasterski:2020xvn,Pasterski:2021dqe}.
At integer points the representation $D_{\chi}$ has an invariant subspace and so is \textit{reducible}. For example suppose $n_1$ and $n_2$ are both positive integers. Then $D_{\chi}$ contains polynomials of the form 
\be
\phi(z,\bar z) = \sum_{i=0}^{n_1 -1} \sum_{j=0}^{n_2-1} c_{ij} z^i \bar z^j,
\ee
which are closed under the action of SL$(2,\mathbb C)$. This invariant subspace is called $E_{\chi}$. This has dimension $n_1 n_2$. 

Now consider the `shadow' representation $D_{-\chi}$. Within this representation there is a invariant subspace which can be described as follows. Consider the space of all functions $\phi(z,\bar z)\in D_{-\chi}$ which satisfy the conditions 
\be
b_{ij} = \int dz d\bar z z^i \bar z^j \phi(z,\bar z) = 0 , \   0\le i \le n_1-1, \  0 \le j \le n_2 -1.
\ee
One can show that under the SL$(2,\mathbb C)$ action on $\phi(z,\bar z)$ the numbers $b_{ij}$ transform linearly and so the space described by the conditions $b_{ij}=0$ is invariant under the action of SL$(2,\mathbb C)$. 
This invariant subspace of $D_{-\chi}$ is denoted by $F_{-\chi}$.  

We can phrase this more cleanly as follows. Comparing to our discussion of generalized functions above we see that the residue of the shadow kernel at these integer points reduces to a differential operator \be\label{difdesc}
\p^{n_1}\bar\p^{n_2}: D_{\chi}\mapsto D_{-\chi}.
\ee
The subspace $E_\chi\subset D_{\chi}$ is the kernel of this operator. Meanwhile $f\in F_{-\chi}$ can be written as $\p^{n_1}\bar\p^{n_2} d$  for some $d\in D_{\chi}$. Namely
\be
\p^{n_1}\bar\p^{n_2}D_{\chi}=F_{-\chi}\subset D_{-\chi}.
\ee
To construct a non-degenerate pairing with elements of $E_{\chi}$ we need to restrict to the equivalence class $[D_{-\chi}]\sim D_{-\chi}+F_{-\chi}$ which is the co-kernel of our differential map~\eqref{difdesc}. This is closely related to the equivalence classes encountered in the BMS flux algebra of~\cite{Barnich:2021dta,Donnay:2021wrk}, however for the leading through sub-subleading radiative soft gravitons only one of the two (holomorphic or anti-holomorphic) submodules has a primary descendant.

\section{Complexifying the Celestial Sphere}\label{app:cpx_sphere}

We will use this appendix to tie together some loose ends regarding the contour prescriptions for our modified shadow transform. Our starting point will be to view the complexified Riemann sphere as the cross section of the complexified null cone in momentum space. Let $p^\mu \in \mathbb{C}^4$ be coordinates on complexified momentum space.  The complexified celestial sphere is then the following quadric in $\mathbb{CP}^3$
\be
-(p^0)^2+(p^1)^2+(p^2)^2+(p^3)^2=0.
\ee
This is known to be bi-holomorphic to $\mathbb{CP}^1\times \mathbb{CP}^1$ via the Segre embedding $\mathbb{CP}^1\times \mathbb{CP}^1\rightarrow\mathbb{CP}^3$
\be\label{eq:segre}
([x:y],[a:b])\mapsto ([xa:ya:xb:yb])
\ee
up to a linear transformation. 
In terms of the standard coordinates for covering the north pole patch of the celestial sphere
\be
[x:y]=[1:z],~~~~[a:b]=[1:\bz]
\ee
the map~\eqref{eq:segre} reduces to
\be\label{zbz}
[x^0:x^1:x^2:x^3]=[1:z:\bz:z\bz]=[p^0+p^3,p^1+ip^2,p^1-ip^2:p^0-p^3].
\ee
We thus see that the complexified celestial sphere can be naturally though of as $\mathbb{CP}^1\times \mathbb{CP}^1$.  The celestial sphere in $\mathbb{R}^{1,3}$ corresponds to the locus $z^*=\bz$, while the locus $\{z^*=z, \bz^*=\bz\}$ lands us on (a quotient of) the celestial torus, relevant to $\mathbb{R}^{2,2}$.

As an example of how this complexification of the celestial sphere is useful, let us examine how a different choice of contour prescription allows~\cite{Donnay:2021wrk} to evade the obstruction encountered by~\cite{Fotopoulos:2019vac}. We wil start with the $\Delta=3,J=\pm1$ modes sourcing the subleading soft graviton 
\be
\J=\J_{soft}+\J_{hard},~~~\bar\J=\bar\J_{soft}+\bar\J_{hard},
\ee
where, at linearized order,
\be
\J_{soft}=-\frac{1}{2}\bar \p^3 S,~~~ \bar \J_{soft}=-\frac{1}{2} \p^3 \bar S.
\ee
The authors of~\cite{Donnay:2021wrk} define the celestial stress tensor as follows
\be
{\bf T}(z)=\oint_C\frac{d\bz}{2\pi i} \bar\J_{soft}(z,\bz),~~~\bar {\bf T}(\bz)=\oint_C\frac{dz}{2\pi i} \J_{soft}(z,\bz).
\ee
If we formally evaluate the shadow transform~\eqref{shadow} using the contour prescription (see also the recent discussion in~ \cite{He:2022zcf})
\be\label{new_contour}
\int d^2 w\Rightarrow \pi \oint \frac{dw}{2\pi i}\oint\frac{d\bw}{2\pi i}
\ee
of~\cite{Donnay:2021wrk} we get $\bar T(z,\bz)\Rightarrow \bar{\bf T}(\bz)$ which is anti-meromorphic by construction. Now in Appendix~\ref{app:gelfand} we saw that our understanding of generalized functions is tied to the parings of the form~\eqref{pairing} which involve a choice of metric. For example, a double contour of the form~\eqref{new_contour} gives
\be
\lim_{\alpha\rightarrow 0}\alpha \oint \frac{dz}{2\pi i}z^{-1+\alpha}\oint\frac{d\bz}{2\pi i}\bz^{-1+\alpha}=0
\ee
in contrast to~\eqref{d2z}. This is consistent with the fact that the Green's function for $\p_z^{n_1} \p_\bz^{n_2}$ is signature-dependent.  For example when we restrict to the codimension 2 locus $(z,\bz)\in\mathbb{R}^{1,1}$ rather than $\p_\bz z^{-1}=\pi\delta(z,\bz)$ we get  $\p_\bz [\theta(\bz)\delta(z)]=\delta(z)\delta(\bz)$ on the locus $\bz=z^*$. The upshot is that (so long as we analytically continue our dimensions slightly away from the integer points) we can start from the amplitudes, complexify our mode expansion so that $(z,\bz)\in\mathbb{C}^2$, take derivatives in $z$ and $\bz$ as we see fit, but when it comes to discussing what contact terms appear our choice of contour matters.

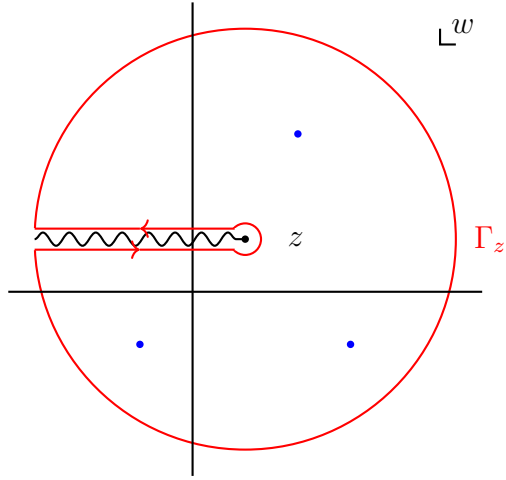
\begin{figure}[th!]
\begin{center}
\begin{tikzpicture}[scale=.7]
\draw[thick] (5,5-.3)--(5-.3,5-.3) node[above right]{$w$} --(5-.3,5);
\draw[thick,red] (1,1) circle (3mm);
\draw[thick,red] (1,1) circle (40mm) node[right]{$~~~~~~~~~~~~~~~~~~~~~\Gamma_z$};
\draw[white, fill=white] (-3-.2,1-.2) -- (1,1-.2) -- (1,1+.2) -- (-3-.2,1+.2) -- (-3-.2,1-.2);
\draw[thick, fill]  (1,1) circle (.5mm) node[right]{$~~~z$};
\draw[thick,branch cut] (-3,1) -- (1,1);
\draw[thick] (-3.5,0) --(5.5,0);
\draw[thick] (0,-3.5) --(0,5.5);
\draw[thick,->,red] (-3,1-.2) -- (-1,1-.2);
\draw[thick,red] (-1,1-.2) -- (1-.2,1-.2);
\draw[thick,red] (-3,1+.2) -- (-1,1+.2);
\draw[thick,<-,red] (-1,1+.2) -- (1-.2,1+.2);
\draw[thick,fill,blue] (2,3) circle (.5mm);
\draw[thick,fill,blue] (3,-1) circle (.5mm);
\draw[thick,fill,blue] (-1,-1) circle (.5mm);
\end{tikzpicture}
\end{center}
\caption{Branch cut for the shadow kernel in the complexified celestial sphere coordinate $w$. The blue dots indicate locations of other operator insertions which from the form of the tree level celestial OPEs are known to give poles in $w$. 
\label{fig:branch}
}
\end{figure}

For our discussions in the main text, it's important to note that the shadow integral kernel has a branch cut once $z$ and $\bz$ are independent. Indeed, the Green's function we are using in section~\ref{sec:newshad} can be rephrased in terms of Cauchy's residue theorem.  Consider a function $f(w)$ that is holomorphic near the point $w=z$. By Cauchy's integral formula the $n$th derivative at $w=z$ can be extracted by the contour integral
\be\label{cauchy_int}
f^{(n)}(z)=\frac{\Gamma(n+1)}{2\pi i}\oint_{C_z}dw\frac{f(w)}{(w-z)^{n+1}},
\ee
where the contour $C_z$ is along a closed curve going counterclockwise around the point $w=z$. If we try to analytically continue $n=1-2h$ away from an integer value, the integrand develops a branch cut stretching between $w=z$ and $w=\infty$.  We can form a closed contour by replacing $C_z$ with a keyhole contour $\Gamma_z$.  Because $f(w)=\phi(w)\sim w^{-2h}$, the integrand goes like $w^{-2}$ so there is no residue at infinity.  The contribution from two points on opposite sides of the branch cut in Figure~\ref{fig:branch} will differ by the following phase
\be
e^{\pi i(n+1)}-e^{-\pi i(n+1)}=-2i \sin (n\pi)=\frac{2\pi i}{\Gamma(n+1)\Gamma(-n)}
\ee
so that 
\be
f^{(n)}(z)\ni \frac{1}{\Gamma(-n)}\int_{-\infty}^z\frac{f(w)}{(w-z)^{n+1}}
\ee
matching~\eqref{greensfn}.
The small arc and branch cut contributions are related by the residues that appear from collinear limits with other operator insertions in a given $\cal S$-matrix element (indicated schematically by the blue dots in figure~\ref{fig:branch}).

\section{\texorpdfstring{OPE between modified stress tensors: $\bar T_{mod}\bar T_{mod}$}{OPE between modified stress tensor}}\label{app:ope-mod} 
Let's start by reviewing the elements of the derivation in~\cite{Fotopoulos:2019vac} that get modified once we take into account the reparameterization mode.  In order to reproduce the standard centerless Virasoro symmetry, a certain smearing of the subleading soft mode must be shown to vanish 
\be
\label{Tint}
\begin{gathered}
\langle{\bar T(z,\bz) \bar T(w,\bw)\prod_{i=2}^n \phi_{h_i,\bar h_i}(z_i,\bar z_i)}\rangle 
= \text{Standard terms with $c=0$ } ~~~~~\\ 
~~~~+ \frac{6}{\pi(\bar w -\bar z)^2} \int d^2 z_1 \frac{1}{(\bar z - \bar z_1)^2 (\bar w -\bar z_1)^2} \langle{S(z_1,\bar z_1)\prod_{i=2}^n \phi_{h_i,\bar h_i}(z_i,\bar z_i)}\rangle.
\end{gathered}
\ee
One needs to be careful, since while the angular dependence of the regulated $z_1$ integral vanishes due to conformal invariance (c.f. section 4.1 of~\cite{Fotopoulos:2019vac}) the   in the regulated $z_1$, there is an overall factor of $\Gamma[0]$.  One thus needs to be careful not to drop subleading terms in the $\Delta\rightarrow0$ expansion of the object it is multiplying. By taking a $w$ derivative of both sides, we can see that there is an obstruction to this integral vanishing via~\eqref{eq:bottomdiamond} and~\eqref{Sdec}. 

We can now show that if we repeat the same manipulations that led to the obstruction~\eqref{TdecT} with $\bar T$ replaced with $\bar T_{\rm mod}$, the non-contact terms vanish.  The correlation function we are interested in takes the form
\be\scalemath{1}{
\badat{3}
&\langle{\bar T_{mod}(z,\bar z) \bar T_{mod}(w,\bar w) \prod_i \phi_i(z_i,\bar z_i)}\rangle =\\
&~~~~~~~~~~~~~~~\langle{\bar T(z,\bar z) \bar T(w,\bar w) \prod_i \phi_i(z_i,\bar z_i)}\rangle +\langle{\frac{1}{2}\bar\partial^3 \epsilon_{S}(z,\bar z) \frac{1}{2}\bar\partial^3 \epsilon_{S}(w,\bar w) \prod_i \phi_i(z_i,\bar z_i)}\rangle\\
&~~~~~~~~~~~+  \langle{\bar T (z,\bar z) \frac{1}{2}\bar\partial^3 \epsilon_{S}(w,\bar w) \prod_i \phi_i(z_i,\bar z_i)}\rangle 
+ \langle{\frac{1}{2}\bar\partial^3 \epsilon_{S}(z,\bar z) \bar T (w,\bar w)  \prod_i \phi_i(z_i,\bar z_i)}\rangle.
\eadat}
\label{eq:tmodtmod}
\ee
Now, as discussed in section~\ref{sec:subsoft}, the first term is~\eqref{Tint}. Using that \be
\label{epsilon1}
\langle{\bar\partial^3 \epsilon_{S}(z,\bar z) \prod_i \phi_i(z_i,\bar z_i)}\rangle = 0
\ee
as an operator equation, and that we can choose the memory operators~\eqref{eq:eps} to have vanishing correlators~\cite{Pasterski:2021dqe}
\be
\langle\frac{1}{2}\bar\partial^3 \epsilon_{S}(z,\bar z) \frac{1}{2}\bar\partial^3 \epsilon_{S}(w,\bar w)\rangle = 0
\ee
we see that the second term on the right hand side of~\eqref{eq:tmodtmod} vanishes. To obtain the mixed correlators involving $\bar T$ and $\epsilon_S$  we use the relation
\be\scalemath{.95}{
\begin{gathered}
\langle{\bar T(z,\bar z) \bar\partial^3 S(w,\bar w) \prod_{i=1}^n \phi_{h_i,\bar h_i}(z_i,\bar z_i)}\rangle 
= \( - \frac{24}{\(\bar z - \bar w\)^5} - \frac{12}{\(\bar z - \bar w\)^4} \frac{\partial}{\partial \bar w}\)\langle{S(w,\bar w)\prod_{i=1}^n \phi_{h_i,\bar h_i}(z_i,\bar z_i)}\rangle
\end{gathered}}
\ee
and the fact that $S(z,\bar z)$ and $\epsilon_{S}(z,\bar z)$ transform in the same way under $\overline{Vir}$ transformations. This follows from the definition $S = \partial\epsilon_{S}$ and leads to 
\be
\label{epsilon}\scalemath{0.95}{
\begin{gathered}
\langle{\bar T (z,\bar z) \frac{1}{2}\bar\partial^3 \epsilon_{S}(w,\bar w) \prod_i \phi_i(z_i,\bar z_i)}\rangle 
= \( - \frac{12}{\(\bar z - \bar w\)^5} - \frac{6}{\(\bar z - \bar w\)^4} \frac{\partial}{\partial \bar w}\)\langle{\epsilon_{S}(w,\bar w)\prod_{i=1}^n \phi_{h_i,\bar h_i}(z_i,\bar z_i)}\rangle,
\end{gathered} }
\ee
where we have also used~\eqref{epsilon1}. Putting everything together we get the standard answer with zero central charge
\be
\begin{gathered}
\langle \bar T_{mod}(z,\bar z) \bar T_{mod}(w,\bar w) \prod_i \phi_i(z_i,\bar z_i) \rangle = \\
\( \frac{2}{(\bar z - \bar w)^2} + \frac{1}{\bar z - \bar w}\frac{\partial}{\partial\bar w}\)\langle{ \bar T_{mod}(w,\bar w) \prod_i \phi_i(z_i,\bar z_i)}\rangle\\ 
+ \sum_i \( \frac{\bar h_i}{(\bar z - \bar z_i)^2} + \frac{1}{\bar z - \bar z_i}\frac{\partial}{\partial\bar z_i}\)\langle{\bar T_{mod}(w,\bar w) \prod_i \phi_i(z_i,\bar z_i)}\rangle.
\end{gathered}
\ee

\bibliographystyle{utphys}
\bibliography{references}

\end{document}